\documentclass[]{spie}  

\newcommand{\rhos}{\rho_\odot}
\newcommand{\rhop}{\rho_\oplus}
 
\usepackage{amsmath,amsfonts,amssymb}
\usepackage{graphicx}
\usepackage[colorlinks=true, allcolors=blue]{hyperref}

\title{RISTRETTO: the PIAA Nuller in the prototyping phase}

\begin{document}

\authorinfo{Further author information, send correspondence to N. Restori (nathanael.restori@unige.ch)}

\author[a]{N. Restori}
\author[a]{N. Blind}
\author[b]{J. K\"uhn}
\author[a]{B. Chazelas}
\author[a]{C. Lovis}
\author[c]{C. Mordasini}
\author[a]{M. Shinde}
\author[d]{P. Martinez}
\author[e,f,g]{O. Guyon}

\affil[a]{Observatoire astronomique de l'Universit\'e de Gen\`eve, Versoix, Switzerland}
\affil[b]{Space Sciences Institute, University of Bern, Bern, Switzerland}
\affil[c]{Weltraumforschung und Planetologie, Physikalisches Institut, Universität Bern, Gesellschaftsstrasse 6, CH-3012 Bern, Switzerland}
\affil[d]{Observatoire de la C\^ote d’Azur, CNRS, Laboratoire Lagrange, Nice, France}
\affil[e]{University of Arizona, Steward Observatory, Tucson, Arizona, United States}
\affil[f]{National Astronomical Observatory of Japan, Subaru Telescope, National Institutes of Natural Sciences, Hilo, HI96720, USA}
\affil[g]{Astrobiology Center, National Institutes of Natural Sciences, Osawa, Mitaka, Tokyo, JAPAN}

\maketitle

\begin{abstract}
The objective of the coronagraphic IFU of RISTRETTO is to enable High Dispersion Coronagraphy of planets at a distance of 2$\lambda$/D from their star, without compromising on transmission. The new idea of a PIAA Nuller (PIAAN) allows contrast down to 10$^{-5}$ over large bandwidth $\ge$ 25\%, with high transmission $\ge$ 70\% at the distance of 2$\lambda$/D. While RISTRETTO will be installed on a VLT, this development is of tremendous importance for fully exploiting future ELTs XAO. We will discuss our PIAAN prototyping activities. This covers 1) the characterisation of our 2nd set of IFU bundles, with 3D-printed MLAs; 2) the characterisation of our first PIAA optics;  3) the integration of a high contrast bench, planned for prototyping of Front-End control strategies; 4) the characterisation of the PIAAN system on the bench.
\end{abstract}


\section{INTRODUCTION}
\label{sec:intro}

Direct detection of the light from extrasolar planets is a challenging objective. For young exoplanets on wide orbits this is achieved by high-contrast imaging and coronagraphy. For close-in exoplanets it is possible to take advantage of planetary transits and obtain a direct measurement of the IR flux of giant planets using secondary eclipses and phase curves. However, many known exoplanets, including those orbiting the brightest and nearest stars, remain out of reach of these techniques.

The RISTRETTO instrument will attempt reflected-light observations of spatially-resolved exoplanets for the first time. The proposed technique for RISTRETTO is High Dispersion Coronagraphy (HDC) spectroscopy, that can enable the $10^{-7}$ contrast level. The Front-End alone must achieve a contrast level of 10$^{-4}$. This will be achieved thanks to a new type of coronagraph, so-called PIAA-Nuller (PIAAN), supported by an eXtreme AO to reduce atmospheric turbulence and telescope disturbances. The XAO is described in these proceedings \cite{blind_2024a}. Using a high-resolution spectrograph \cite{chazelas_2024a}, the radial velocity shift between star and planet allows us to separate the stellar and planetary spectral lines, providing an additional factor $\sim$1000 in achievable contrast. Driven by the Proxima b science case, the RISTRETTO instrument will be targeting reflected light in the visible wavelength range. The short wavelengths allow us to spatially resolve Proxima b and other known very nearby exoplanets, placing them at about 2 $\lambda/D$ on a 8m-class telescope.

In the case of a photon-noise limited measurement, the SNR of HDC technique is given by \cite{lovis_2016a}:
\begin{equation}
    \mathrm{SNR} \varpropto \sqrt{T} \dfrac{\rhop}{\sqrt{\rhos}},
\end{equation}
where $T$ is the total transmission, and $\rhop$ and $\rhos$ are respectively the planet and the stellar light coupled into the external fibers of the IFU. Performance of RISTRETTO rely on 2 fundamentals requirements:
\begin{enumerate}
    \item $\rhop \ge 50\%$ for host stars as faint as PDS70 (Imag = 11).
    \item $\rhos \le 10^{-4}$ for Proxima Cen.
\end{enumerate}
, applying for median seeing conditions (Seeing = 0.83" at zenith distance of $30^\circ$, $L_0$ = 20m, Wind = 9.5m/s). Note that maximum elevation of Proxima Cen is 52deg, i.e. a corrected median seeing $\sim$0.90”.

This paper focuses on the current PIAA-Nuller prototyping activities, with the recent delivery of a 2nd IFU as well as our 1st PIAA optics.

\begin{figure}[b]
    \centering
    \includegraphics[width=1\textwidth, trim={0 0 26cm 0}]{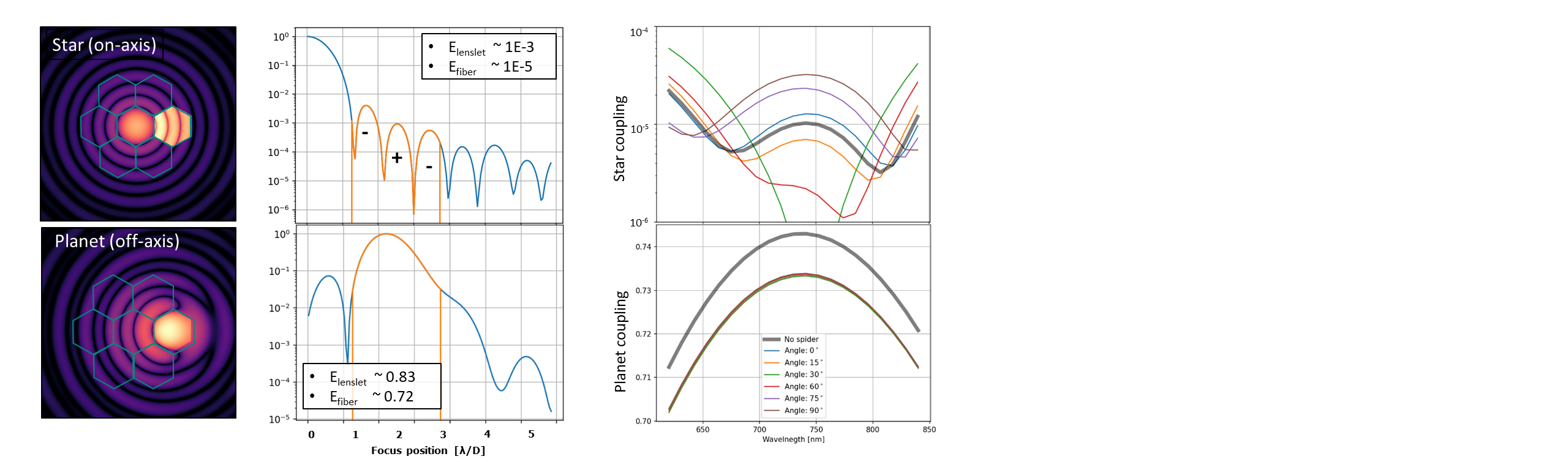}
    \caption{Working principle of the PIAA Nuller. Top: coupling of the star (on-axis). Bottom: coupling of the planet at 2$\lambda/D$. Left: PIAA PSF and overlayed IFU at 620nm, with the 'planet' lenslet showcased. Middle: Horizontal cut view of the PSF in log scale, with yellow part showing the position of the planet lenslet. Right: Expected stellar and planet coupling on the planet lenslet between 620 and 840nm. Each color represents a different fiber, with differences due to the effect of spider diffraction.}
    \label{fig:piaan_concept}
\end{figure}

\section{The PIAAN}
\label{sec:piaan}

We remind here the working principle. More details can be found in \cite{blind_2022a}. The PIAAN combines a PIAA optics with an IFU, made of 7 singlemode fibers arranged in an hexagonal array. The PIAA optics partially apodizes the pupil, and doing so, can provide some level of nulling of the star over the external SMFs of the IFU. This solution leads to a win-win situation, where $\rhos$ is optimized and $\rhop$ is minimized.

\begin{itemize}
    \item For \textbf{stellar coupling} (Fig.~\ref{fig:piaan_concept}, top), we follow the idea of a nuller by partially apodizing the pupil, hence generating diffraction rings. Those rings are about 10 to 40 times fainter than in the unapodized case, and have higher spatial frequency, so that 2 (at $\lambda=840$nm) to 4  (at $\lambda=620$nm) are present in a lenslet. Because the electric field of those rings is changing from one to the other, they eventually cancel each other over the SMF mode, hence creating nulls to the $10^{-5} - 10^{-6}$ levels.
    \item \textbf{Planet coupling} (Fig.~\ref{fig:piaan_concept}, bottom) is improved by reducing energy contained in the diffraction rings, and which would be filtered out by the lenslet otherwise. For the current designs, we generally get about 80-85\% of the flux into the lenslet at 2$\lambda/D$. The PIAA off-axis aberrations are reasonable thanks to the moderate apodization.
    The electric field transmitted by the lenslet is still a rather good match to the SMF mode, so that the coupling from lenslet to fiber is around 90\%. This leads to $\rhop^0 \sim 70-75\%$ for the best solutions, slightly higher than with a VLT pupil with similar IFU. Considering AO residuals, we can also deduce that $\rhop = \rhop^0 \times \mathcal{S}$, where $\mathcal{S}$ is the Strehl ratio.
\end{itemize}
This makes our PIAA design one of the highest transmission coronagraph with inner working angle of 1.5 to 2.5$\lambda/D$ and band-pass of $\sim 30\%$. The higher transmission of the PIAA is a fundamental advantage once on sky, since solutions with lower transmission lead to more stringent constraints on the Strehl, i.e. the actuator count and/or seeing conditions. 

The PIAAN, like any coronagraph with such small inner working angle, is extremely sensitive to low order aberrations. For all our design and tolerancing work, we use the following tolerance: \textbf{WFE($f \le$ 6 cycles) $\le$ 10 nm RMS}.

\begin{figure}[t]
    \centering
    \includegraphics[width=0.9\textwidth, trim={0 0 12cm 0}]{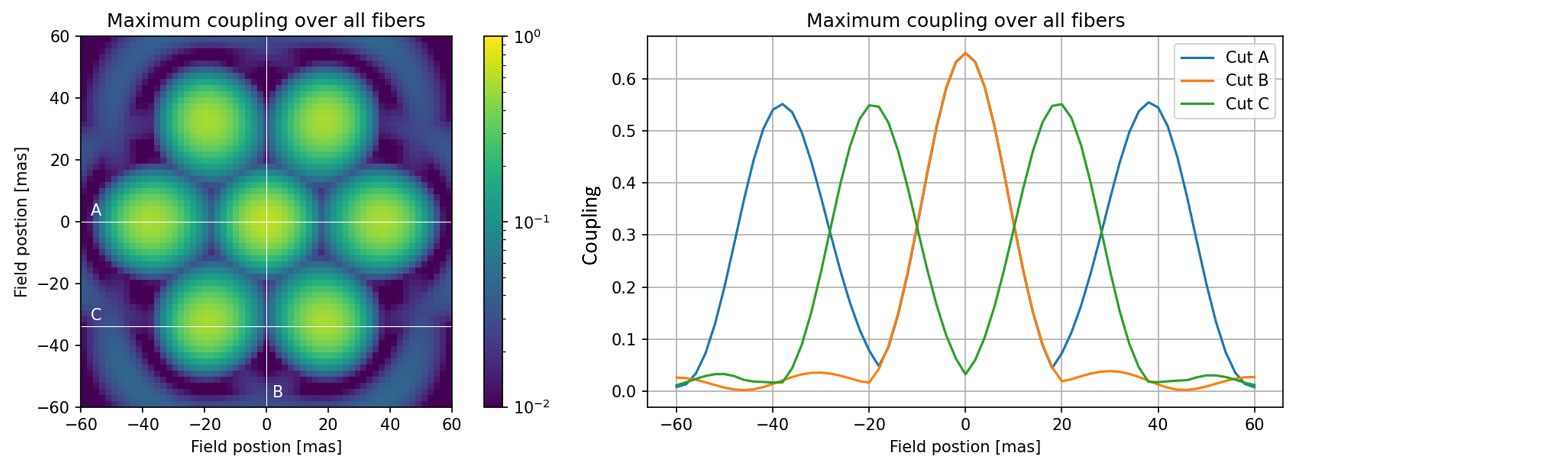}
    \caption{Left: Transmission of the PIAAN over the 7 fibers as a function of the object position. Right: 3 cuts through the map showing the fast loss of transmission if the planet is not centered, and in particular the blind zones between fibers with transmission dropping to zero (cut B \& C).}
    \label{fig:piaan_throughput}
\end{figure}

Its field of view is also limited by the nature of singlemode fibers (Fig.~\ref{fig:piaan_throughput}). A lenslet transmission drops quickly if the object is not centered, with 50\% loss at $\pm$ 0.5$\lambda/D$ ($\pm$ 9mas for RISTRETTO). If the planet falls between 2 fibers, transmission drops nearly to 0 (Fig.~\ref{fig:piaan_throughput}, cuts B \& C). Since the position of the planet is (a priori) unknown, we plan to make two exposures with the IFU rotated by 30$^\circ$ to balance its transmission over the FoV. This can be accounted as a transmission loss of 50\% due to the doubled integration time. The same argument and strategy holds in case of a spaxel with lower performance: a spaxel with 0 transmission would account for 1/6 transmission loss with proper observing strategy). There is therefore a strong interest in determining where the planet is by other means, to place it directly on the best spaxel, and even optimize the PIAAN working point for it.

\paragraph{Additional developments}
By design, the PIAAN is not very robust to asymmetries in the pupil plane, starting with telescope spiders. The thin VLT spiders do not have a strong impact on contrast, although the turbulence-free figure becomes more complex (Fig.\ref{fig:ifu_perf}). This will complexify the determination of a working point due to the differential field-pupil rotation. 

Some preliminary work looks at symmetrizing the PIAA PSF over the IFU. This is based on the idea of the SCAR coronagraph\cite{por_2018b}, a pupil plane coronagraph that distorts the PSF to create nulls over an IFU similar to RISTRETTO's. In our case, 90\% of that nulling work should be performed by the PIAA, so a slight modification of the pupil (in intensity and/or phase) should balance the behavior of the 6 external spaxels with a limited transmission cost. This work could also benefit to finding the optimal working point of the PIAAN. Until we characterize a complete PIAAN prototype and know where we stand also with respect to the XAO limitation, this will have low priority.

\section{The Integral Field Unit}
\label{rsec:ifu}

Since last SPIE\cite{kuhn_2022a}, a different prototype IFU was produced. The 1st one showed encouraging results, despite 'historical' choices that limited its performance. The 2nd prototype delivered early 2024 shows performance that could almost qualify it as a final version.

The IFU unit consists in a bundle of 7 standard SMF, packed in an hexagonal array. A micro-lens array is then 3D-printed on top of the bundle.

\subsection{The singlemode fibers}

We use off-the-shelf Thorlabs SM630HP singlemode fibers. They were thoroughly characterized, in particular regarding the NA which must be known to better than 5\% to optimize the MLA design. 

\begin{figure}[t]
    \centering
    \includegraphics[width=0.7\textwidth, trim={0 0 16cm 0}, clip]{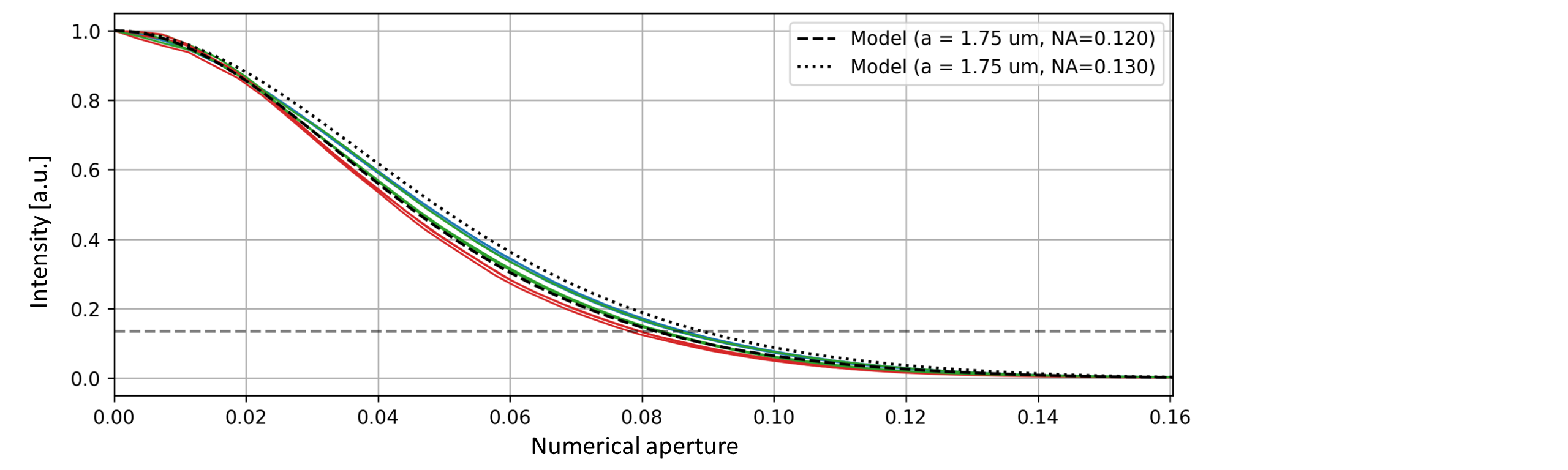}
    \caption{Thorlabs SM630HP NA characterisation for different fibers (color lines). SMF models in dashed and dotted black lines.}
    \label{fig:smf_charac}
\end{figure}

We directly image their far-field on a CCD, without optics. We acquire images at different CCD-fiber distances, which allows to extract the exact distance for each step and measure the angular scale of the setup. The LP$_{01}$ mode is then fitted to the data. We estimated NA(1/$e^2$) = 0.083 ($\pm 3\%$) (or geometrical NA=0.120), for several of those fibers coming from Thorlabs (bare or connectorized) or from a sample provided by SQS/AMS (Fig.~\ref{fig:smf_charac}).

\subsection{Bundle arrays}
\label{sec:bundle}

In 2024, two bundle prototypes were produced by AMS/SQS with a pitch of 125$\mu m$ (Fig.~\ref{fig:ifu_proto}, A). For these prototypes, we use an 'octopus' configuration, with 7 FC/PC fibers as output of the hexagonal bundle. For the final bundle, the 7 fibers will be grouped in a linear slit at the output.

The two bundles are well within our specifications regarding core-to-core pitch, far-field co-alignment (i.e. mechanical parallelism of fibres), as well as perpendicularity to the mechanical surface  (Tab.~\ref{tab:ifu_perf}). 

\begin{table}[h]
        \centering
    \caption{SQS bundle arrays performance.}
    \vspace{0.3cm}
    \begin{tabular}{l|cc|cc}
        Bundle & Pitch & Pitch error & FF alignment & Perpendicularity \\
               & [$\mu m$] & [$\mu m$  PTP] & [$^\circ$ PTP] & [$^\circ$] \\
        \hline
         A  & 124.8 & 2.1 & $\le$0.5 &  $\le$0.05 \\
         B  & 125.1 & 1.7 & $\le$0.4 &  $\le$0.03 \\
    \end{tabular}
    \label{tab:ifu_perf}
\end{table}

\subsection{Micro-lens arrays}
\label{sec:mla}

The micro-lens array are made by Keystone Photonics (formerly Vanguard) using the 2-photon polymerisation 3D printing technique. A 1st prototype with a pitch of 250$\mu m$ was tested in 2023: the large pitch however led to a complex fabrication process (high volume to print, multiple printing steps, etc). The IFU showed reduced, but still promising, performance for a 1st try, with coupling up to $\sim$15\% and capability to achieve null down to $10^{-5}$. 

\begin{figure}[t]
    \centering
    \includegraphics[width=1.\textwidth,  trim={3cm 0 2cm 0}]{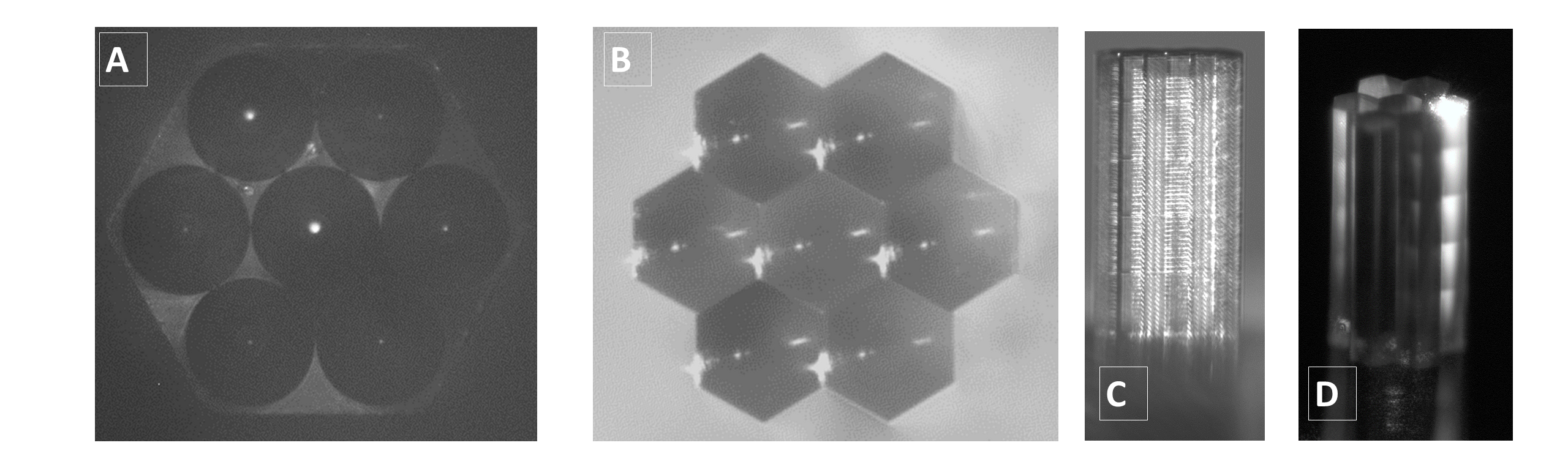}
    \caption{2nd IFU prototype. A) Bundle front face; B) Same view as A) with the MLA printed on it; C) Side view of the MLA, with illumination from the back, showing some of the printing structure. The lenslet height is 0.7726mm. D) Three-quarter view showing the 7-lenslets. A laser source is injected from the fiber to the lenslet for one of them. Diffusion in the lens material shows the beam.}  
    \label{fig:ifu_proto}
\end{figure}

Following that experience, a 2nd prototype is under test, using a minimal pitch of 125$\mu m$ (Sect.~\ref{sec:bundle}; Fig.~\ref{fig:ifu_proto}). Despite the straight pillars design, leakage and coupling from one spaxel to the other seems unlikely, as suggested by the high contrasts we measured (see Sect.~\ref{sec:ifu_perf}).

\begin{figure}[b]
    \centering
    \includegraphics[width=1\textwidth, trim={0 1.5cm 0 0}]{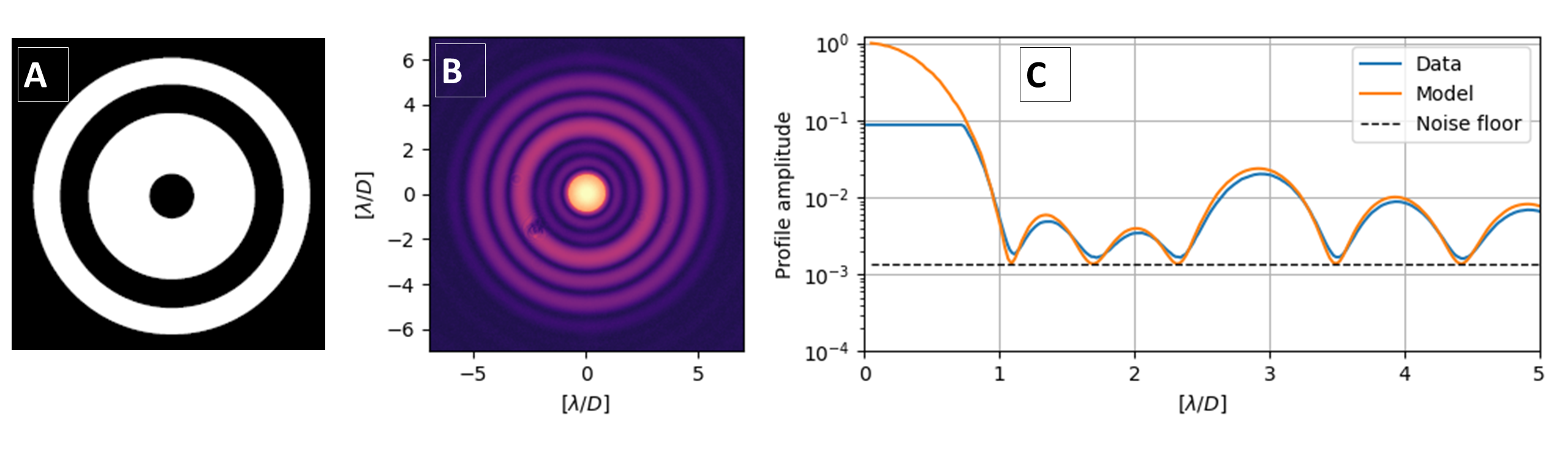}
    \caption{IFU test bench characterisation. A) Modified VLT pupil for achieving nulling without PIAA. B) PSF neasured on the bench at $\lambda$=633nm. C) Integrated profile from (B), compared to the theoretical profile.}
    \label{fig:ifu_bench_charac}
\end{figure}

\subsection{IFU performance}
\label{sec:ifu_perf}

The IFU was tested on a simple bench (different from the high contrast bench presented in Sect.~\ref{sec:high_contrast_bench}), made of two lenses working at f/100, between which is located a ring apodized pupil of diameter 2.54mm\cite{kuhn_2022a} (Fig.~\ref{fig:ifu_bench_charac}). The pupil and bench f-number create a null over the 6 external IFU spaxels.

The MLA is placed at the bench focus on a 5-axis motorized stage to optimize coupling into fibers. A beam splitter is placed in front of the IFU to observe the light back reflected on its surface with a long working distance microscope. Being able to check the alignment of the MLA to the bench as well as verifying proper behavior of the positioners saved us a lot of time.

\begin{figure}[t]
    \centering
    \includegraphics[width=1\textwidth, trim={5cm 0.5cm 7cm 0}]{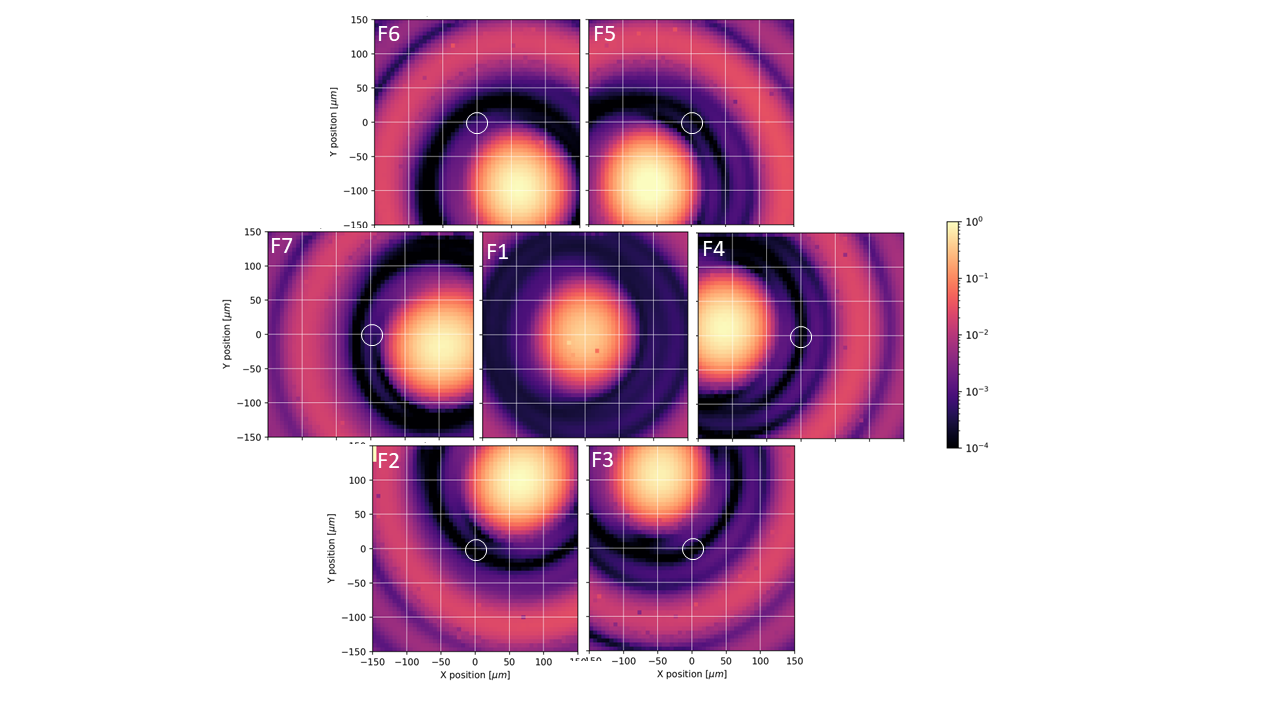}
    \caption{Coupling maps for the 7 fibers of the IFU with the ring apodizer at $\lambda$=633nm. F1 is the central fiber, on which coupling is maximized over 5 axis (optimal (x,y)=(0,0) in the 7 maps). Colormap represents absolute coupling $\rho_{max}^{F_1}$, and contrast $C = \rho^{F_x}/\rho_{max}^{F_1}$ for F2 to F7. }    
    \label{fig:ifu_perf}
\end{figure}

We use two photodiodes to measure flux coupled in two fibers at a time, one being always attached to the central fiber F1. The IFU is first positioned to maximize coupling in F1. We then perform a 2D XY scan, which creates a coupling map for F1 and for F[2-7] (Fig.~\ref{fig:ifu_perf}). The scan helps apprehend the difference in behavior of the 7 spaxels as well as the stability of the bench, since the F1 map is created 6 times (and appears perfectly reproducible to $10^{-5}$ contrast level). But ultimately we are only interested by one question: is there a working point that maximizes contrast for all 6 external fibers? In Fig.~\ref{fig:ifu_perf}, the optimal coupling position of F1 already allows contrasts $C\le2\cdot10^{-4}$ for 4 of the fibers, and $6\cdot10^{-4}$ for F6 and F7 (Tab.~\ref{tab:ifu_perf}). Over all fibers, the deepest measured null reaches $C=5\cdot10^{-6}$. We could distort those maps by applying a low amplitude, low order aberration on a DM to get a more balanced working point/contrast on all 6 fibers, with little cost on transmission. As an illustration with those measurements: a TT offset of (-12,0) $\mu m$ (representing $\sim 0.20 \lambda$/D tip-tilt, or a WFE$\sim \lambda/20$ RMS) balances F[2,3,5,6,7] within specification, at the expanse of F4 and a decrease in planet coupling of 30\%. A more complex aberration (based on a sensitivity analysis of the IFU for instance\cite{blind_2022a, haffert_2020a}) will certainly lead to more balanced performance. This is work we start investigating on the high contrast bench (Sect.~\ref{sec:high_contrast_bench}). 

From those maps, we also extract the best coupling on all fibers, always near the expected position. The apodizer and the MLA design lead to a theoretical maximum of $\rho_{max}^{theo}$=46\%, which is nearly reached with F5. F[1, 2, 4, 6] present 20\% losses, and F[3,7] present 30\% losses (Tab.~\ref{tab:ifu_perf}).

\begin{table}[t]
    \centering
    \caption{Best coupling obtained on the 7 fibers, after optimization on F1. Ratio to maximum should be consider as a transmission factor for RISTRETTO.}
    \vspace{0.3cm}
    \begin{tabular}{l|ccccccc}
        Fiber        & 1 & 2 & 3 & 4 & 5 & 6 & 7\\
        \hline
        $\rho_{max}^{F_x}$ & 0.37  & 0.36 & 0.29 & 0.37 & 0.43 & 0.37 & 0.31 \\
        $T = \rho_{max}^{F_x}/\rho_{max}^{theo}$ & 0.82  & 0.80 & 0.64 & 0.82 & 0.95 & 0.82 & 0.69 \\
        \hline
        $C(0,0) = \rho^{F_x}/\rho^{F_1}$ & -  & $1.1\cdot10^{-4}$ & $0.4\cdot10^{-4}$ & $0.2\cdot10^{-4}$ & $1.5\cdot10^{-4}$ & $6.2\cdot10^{-4}$ & $5.8\cdot10^{-4}$ \\
        $C(-12,0) = \rho^{F_x}/\rho^{F_1}$ & -  & $2.0\cdot10^{-4}$ & $2.3\cdot10^{-4}$ & $13.5\cdot10^{-4}$ & $2.0\cdot10^{-4}$ & $1.0\cdot10^{-4}$ & $0.3\cdot10^{-4}$ \\
        \hline
    \end{tabular}
    \label{tab:ifu_perf}
\end{table}

\newpage
\section{The PIAA optics}
\label{sec:piaa}

Nutek produced our 2 PIAA optics. They were delivered only recently and we report here on preliminary analysis.

Note that instead of having two independent optics, we decided to engrave the two PIAA surfaces in a single CaF2 rod (20mm in thickness). Surface-to-surface centering requirement was well within Nutek capabilities. Due to the air-glass-air configuration, SAGs are more important than for more classical glass-air-glass configuration, but on the other hand, SAG errors have less impact on wavefront error. However, since our PIAA is only a partial apodizer, the SAGs are quite small  ($\sim 70\mu m$) and this solution looked well adapted.

Since we fixed the alignment of the two surfaces, the alignment of the PIAA to the bench is straight forward. Thanks to a peculiar light leakage around the secondary obstruction, alignment to the pupil is also straight forward visually, with a centering accuracy estimated to $\pm$0.5\% of the pupil diameter according to simulations. A pupil mask is planned right in front of the PIAA in practice, in order to cancel some chromatic errors (e.g. pupil chromatic shift due to beamsplitter or ADC).

\begin{figure}
    \centering
    \includegraphics[width=0.8\textwidth, trim={6cm 0cm 8cm 0}]{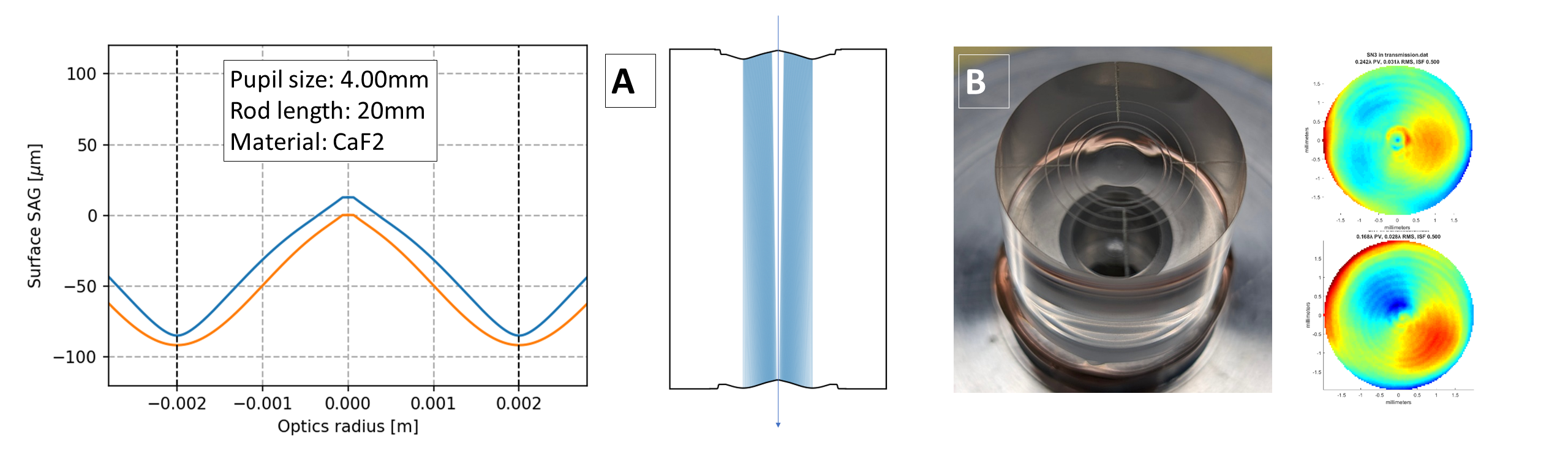}
    \includegraphics[width=1\textwidth, trim={0 0cm 0 0}]{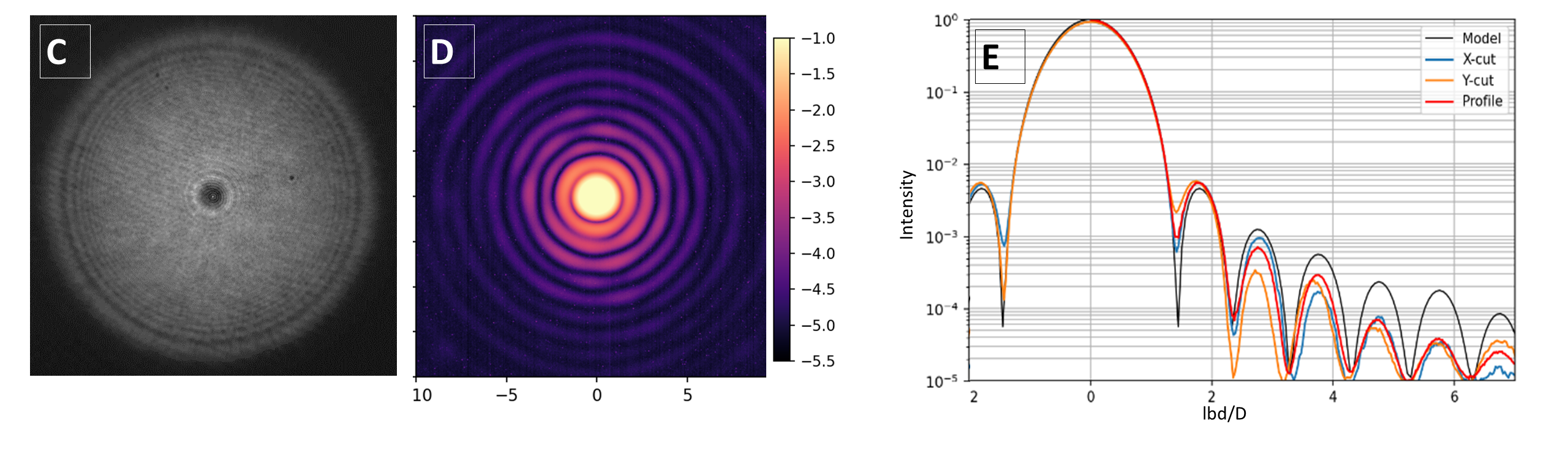}
    \caption{PIAA prototype on the refractive IFU bench of Sect.~\ref{sec:ifu_perf}. A) SAG of the PIAA optics and ray-tracing through the optics (with SAGs exaggerated 10 times). B) Picture of one PIAA optics before coating, and WFE error of both prototypes as measured by Nutek. C) Post-PIAA apodized pupil illumination on bench. D) PIAA PSF in log scale at $\lambda$=633nm. E) Cut views and integrated profile of D) compared to the theoretical PIAA PSF.}    
    \label{fig:piaa_proto}
\end{figure}

The produced optics have a very good optical quality as measured by Nutek on an interferometer (WFE $\le$ 30nm RMS), and as observed on the PIAA PSF (Fig.~\ref{fig:piaa_proto}). The PSF looks more apodized than expected, which could have a negative impact on the contrast performance. We just recently started to characterize them and need to figure this out.

\section{High contrast bench}
\label{sec:high_contrast_bench}

A high contrast bench was built to test in the lab the different components and techniques that will be used in the RISTRETTO Front-End. The bench is built exclusively from off-the-shelf components.

\begin{figure}
    \centering
    \includegraphics[width=0.8\textwidth]{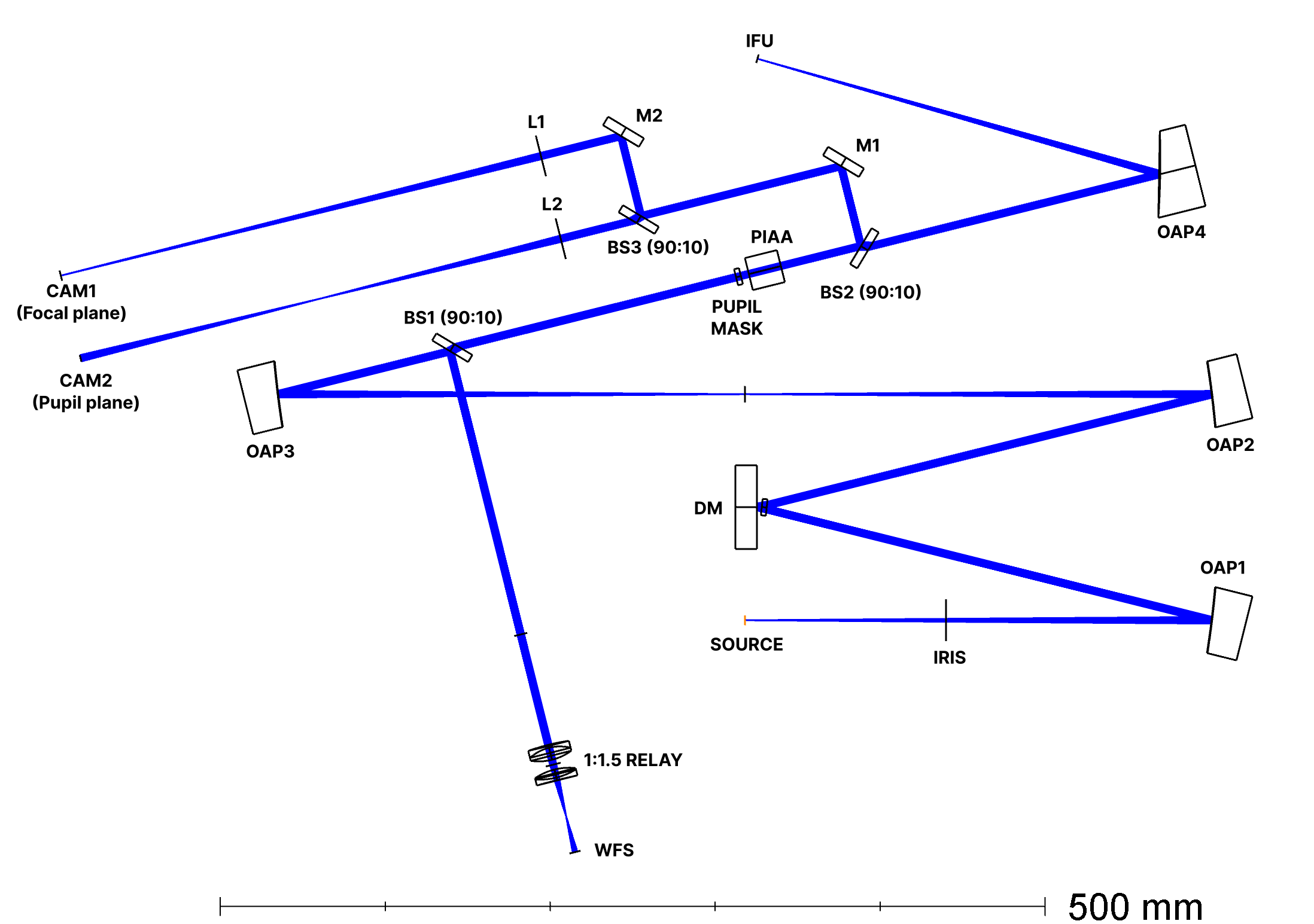}
    \caption{OpticStudio design of the RISTRETTO high contrast bench}
    \label{fig:high_contrast_bench}
\end{figure}

The main path of the bench (Fig.~\ref{fig:high_contrast_bench}) consists of three f=279.10mm OAP from Edmund Optics (OAP1, OAP2, OAP3), a f=254.0mm OAP from Thorlabs (OAP4), a Boston Micromachines Corporation Multi-3.5 DM (12x12 actuators, 3.5$\mu m$ stroke), a chromium pupil mask, and the before-mentioned PIAA. Two cameras (CAM1 and CAM2) allow the visualisation of the PSF and the pupil simultaneously. The IFU plane is motorized using a three-axis translation stage. The optical quality of this (reflective) bench is lower than the (refractive) IFU test bench.

\begin{figure}
    \centering
    \includegraphics[width=0.8\textwidth]{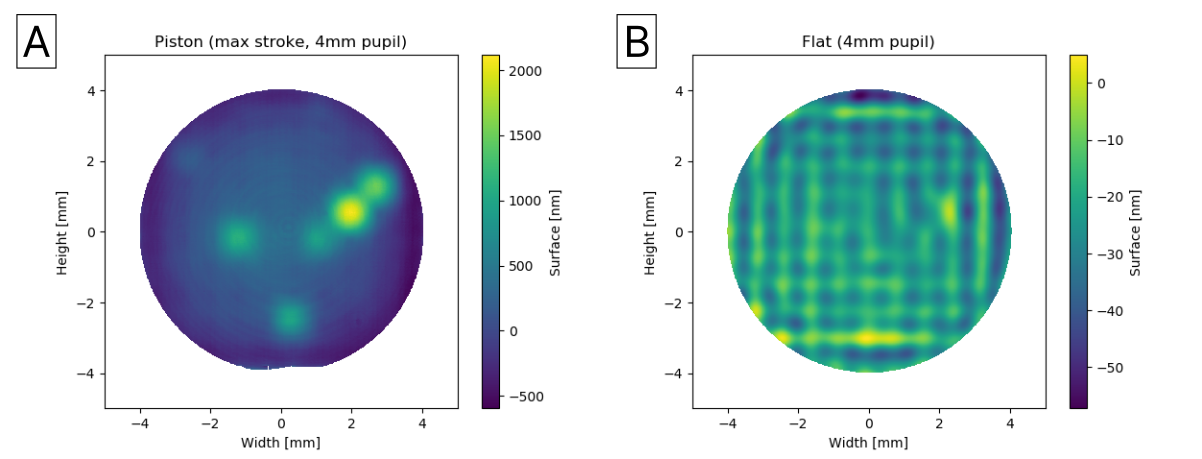}
    \caption{A) Surface map with all actuators at maximum stroke, clearly showing a defocus and the weak actuators. B) Surface map after flattening, showing the residual structure of the actuators them-self.}
    \label{fig:dm_interfero}
\end{figure}

The deformable mirror was first characterized using an interferometer with a 5mm beam  (Fig.~\ref{fig:dm_interfero}). The DM was first used by the KalAO instrument before it was damaged for an unknown reason. The response of each actuator was extracted from individual push-pull measurements, to remove static DM surface contributions. The response matrix and the control matrix of the DM were then computed. Up to six weak actuators were found, with the worst actuator having its stroke reduced to 60\%, but all actuators were functional. As the DM will only be use to make small corrections, the reduced stroke of some actuators was deemed acceptable. The DM was then flattened using a simple integrator loop. The final surface error is 8 nm RMS. A stability measurement of the DM was then done over 24h, with the surface being stable within a 7 nm RMS error, limited by the capabilities and the stability of the interferometer itself.

The pupil of the system is defined by the chromium mask positioned just in front of the PIAA. Different pupils are available: a 4mm pupil without central obstruction and a 4mm VLT pupil without the spiders, and variations of these two pupils with a slightly oversized or undersized primary and secondary. This resulted in aperture diameter varying from 3.6mm to 4mm in 0.08mm steps, and in secondary obstructions with a diameter varying from 0.64mm to 0.78mm in 0.03mm steps, steps being related to the PIAA optics tolerances on the pupil definition. The mask was patterned using photolithography.

Our original intent was to align the bench using an Imagine Optics HASO wavefront sensor, but due to the long focal lengths and the low f\# (f/70 and f/63.5 respectively) of the system, big adjustments were needed to see any changes in the aberrations and the beam was constantly getting out of the field of view of the wavefront sensor, rendering the optical alignment of the bench highly impractical. Mechanical alignment based on the OpticStudio and Solidworks modeling proved itself to be a much more reliable method. The alignment of the PIAA is pretty straightforward, thanks to the monolithic nature of the PIAA and to the two cameras. The PIAA was mounted inside of a four axis gimbal mount, with the lateral alignment ($X-Y$) done by looking at the pupil pattern and the angular alignment ($\Theta X-\Theta Y$) done by looking at the PSF.

\begin{figure}
    \centering
    \includegraphics[width=0.8\textwidth]{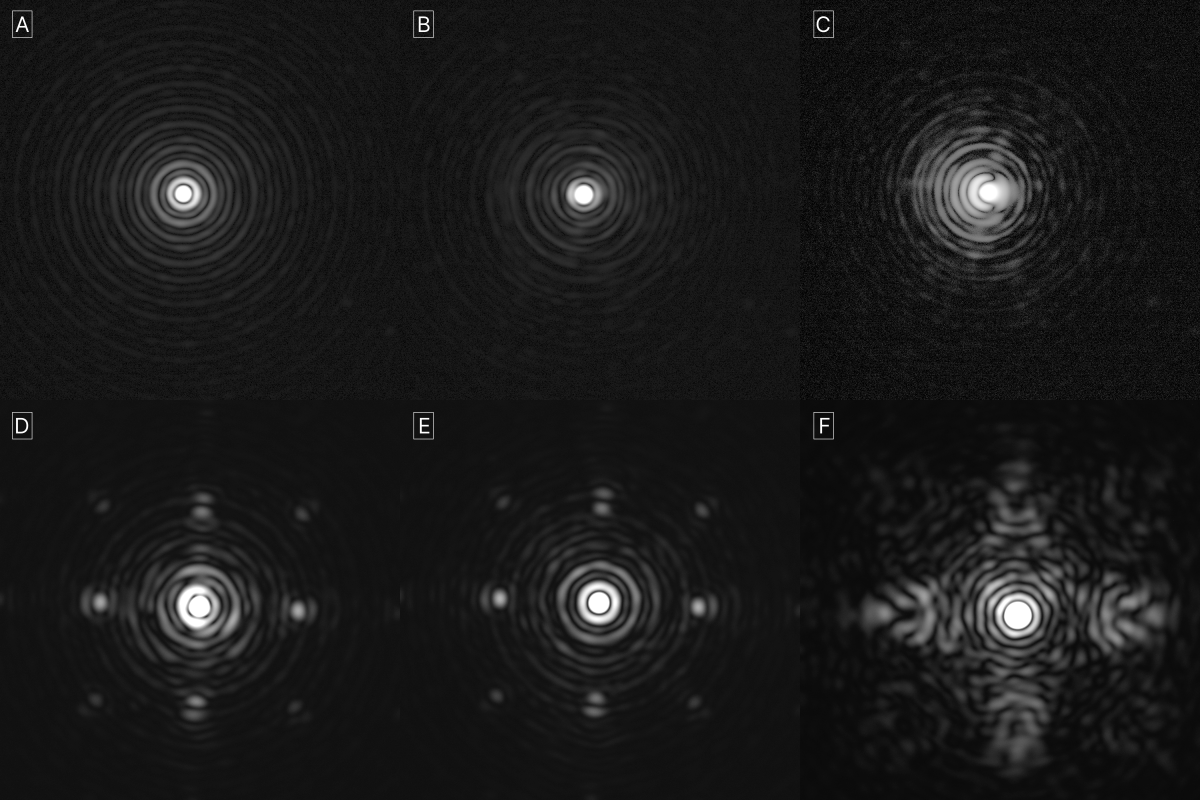}
    \caption{\textbf{With a dummy DM (flat mirror) and a VLT pupil} A) PSF of the bench without PIAA, as aligned. B) On-axis PSF of the bench with PIAA. C) Off-axis PSF of the bench with PIAA. \textbf{With the DM and a VLT pupil} D) PSF of the bench without PIAA, as aligned. Satellite PSFs created by the DM actuators are clearly visible. E) PSF of bench with PIAA, after its optimization, showing clearly separated diffraction rings. F) PSF of bench with PIAA, after its optimization. Satellite PSFs clearly show the shape of the off-axis PSFs of the PIAA.}
    \label{fig:hcb_psfs}
\end{figure}

The PSF of the bench was optimized using the deformable mirror, both without and with the PIAA in place. The eye-doctor technique was used, optimizing the pattern on the DM by scanning and optimizing all modes of a basis one-by-one and doing multiple passes. The figure of merit for the optimization was defined as the average value of a 7x7 pixels region around the peak of the PSF normalized by the total flux on the camera. Two different basis were tried: a Zernike basis and the SVD basis of the DM (computed from the interferometer measurements), with both basis giving similar results. The resulting PSFs can be found in Fig.~\ref{fig:hcb_psfs}. While the PIAA PSF is improved compared to the non-optimized case, the diffraction rings still show some roughness, and in the case of the second and the third ring an asymmetry. The difficulties in optimizing the PIAA PSF may come from the satellite PSFs at about $\pm 5 \lambda/D$ created by the DM actuators: as the off-axis PSFs of the PIAA are quickly spreading with separation from the optical axis of the system, they may interfere with the on-axis PSF rings. A DM with a higher number of actuators might be required to push the satellites further.

It is however difficult to estimate the actual impact on the contrast only from this intensity measurement. Two actions are currently planned around this issue. First, redoing the contrast measurement as described in Sect.~\ref{sec:ifu_perf} on the high contrast bench with the PIAA instead of the nulling mask. This will allow us to assess the actual impact of the ring irregularities and of the satellite PSFs. Secondly, implement a differential-OTF\cite{dotf_martinez} based method for the PSF optimization. As the PIAA is reshaping the pupil, the Zernike modes injected on the DM before the PIAA are not orthogonal anymore after it, potentially leading to issues with the eye-doctor convergence. The dOTF will allow us to make phase measurement after the PIAA, and use a closed loop for the wavefront optimization. The dOTF could actually be used for calibrating the DM influence functions as seen through the PIAA optics, which could help built proper command vectors from the post-PIAA dOTF phase measurements. Ultimately we want to do such a wavefront correction through the IFU itself. We will start with the eye-doctor, but will also study a direct DM-probe technique which, in simulation at least, allows to reconstruct directly the wavefront and command vector at the level of the DM.

A first addition to the bench this summer will be a small fiber-fed prism-based spectrometer. The light from the seven fibers of the IFU bundle will be arranged into a linear slit (similarly to the real spectrograph), with the central fiber attenuated by a fiber-based attenuator. We can then characterize the contrast performance at all science wavelengths (instead of just one laser), and to use that information to study low order wavefront control strategy through the PIAAN, and to balance the contrast in the six peripheral fibers, by introducing low amplitude aberrations (as discussed in Sect~\ref{sec:ifu_perf}). 

The bench is limited to the coronagraphic part for now. We already have extension ideas to make it a form of Front-End prototype and study how all subsystems should work together:
\begin{itemize}
    \item The addition of (residual) turbulence injected via an SLM, obtained from end-to-end simulations. This will make the bench monochromatic for such tests. 
    \item Adding a WFS channel, with pyramid and zernike WFS prototypes.
    \item Adding a closed-loop ADC system.
\end{itemize}

\section{Conclusion}
\label{sec:ccl}

The PIAAN is the first original development of RISTRETTO that will be demonstrated in the lab. Prototyping activities are progressing well, with the delivery of the 2 major sub-components in early 2024. The 2nd IFU prototype is almost at the level of a final version. Several others will be produced in the next 2 years, in particular for the spectrograph AITs and early sky tests. This will be the opportunity to evaluate variations from one run to the other. 

The full PIAAN concept will be tested and characterized by the end of 2024 on the new high contrast bench. We will also start studying the wavefront control strategy through the IFU this year. The goal will be to balance contrast over 6 fibers, or to optimize it on the best one for when we know the planet position.

\section*{Acknowledgement}
This work has been carried out within the framework of the National Centre of Competence in Research PlanetS supported by the Swiss National Science Foundation under grants 51NF40\_182901 and 51NF40\_205606. The RISTRETTO project was partially funded through SNSF FLARE programme for large infrastructures under grants 20FL21\_173604 and 20FL20\_186177. The authors acknowledge the financial support of the SNSF.

\bibliography{references} 
\bibliographystyle{spiebib} 

\end{document}